\documentclass[prb,
twocolumn,
superscriptaddress,showpacs,amsmath,amssymb]{revtex4}
\usepackage{amsfonts}
\usepackage{bm}
\usepackage{verbatim}

\usepackage{graphicx}

\begin{document}

\title{Negative temperature for negative lapse function}

\author{G.E.~Volovik}
\affiliation{Low Temperature Laboratory, Aalto University,  P.O. Box 15100, FI-00076 Aalto, Finland}
\affiliation{Landau Institute for Theoretical Physics, acad. Semyonov av., 1a, 142432,
Chernogolovka, Russia}

\date{\today}

\begin{abstract}
Fermion dynamics distinguishes spacetimes having the same metric $g_{\mu\nu}$, but different 
tetrads $e_{\mu a}$, and in particular, it distinguishes a lapse with negative sign, $N<0$.\cite{Rovelli2012b}
Here we show that the  quasiequilibrium thermodynamic state may exist, in which the region with $N<0$ has negative local temperature $T({\bf r})<0$, while the global Tolman temperature $T_0$ remains positive. For bosons, only $N^2$ matters. However, if bosons are composite,  they may inherit the negative $T({\bf r})$ from the fermions, and thus they may distinguish the spacetimes with positive and negative lapse functions  via thermodynamics. 
 \end{abstract}
\pacs{
}

\maketitle

\section{Introduction}

Tolman's law\cite{Tolman1934} (see e.g. the latest paper on the topic in Ref. \onlinecite{Visser2018}) states that in a static gravitational field (which can be described by the time-independent metric with the shift function $N^i=0$), the locally measured coordinate-dependent temperature $T({\bf r})$ obeys:
\begin{equation}
T({\bf r})= \frac{T_0}{\sqrt{g_{00}({\bf r})}}= \frac{T_0}{|N({\bf r})|} \,,
\label{T}
\end{equation}
where $T_0$ is spatially constant in thermal equilibrium, and  $N$ is the lapse function with
$g_{00}({\bf r})=N^2({\bf r})$.  In the ADM parametrization with $N^i=0$, one has $g_{\mu\nu}dx^\mu dx^\nu=N^2dt^2 - g_{ik}dx^idx^k$.

In the effective gravity emerging for quasirelativistic quasiparticles in superfluids,\cite{Volovik2003}  $T_0$ is the
conventional temperature of the liquid as measured by external observer. It is constant in space in thermal equilibrium. The local  temperature $T({\bf r})$ is measured by the local "internal observer", who uses quasiparticles for measurements. The Tolman law in superfluids is related to the Doppler shift in the same way as the gravitational red shift in general relativity.

Fermions interact with gravity via the  tetrads instead of the metric, $g_{\mu\nu}=\eta^{ab} e_{a\mu} e_{b\nu}$. In terms of tetrads, one has $N^2=g_{00}=  (e_{00})^2= (e^0_0)^{-2}$.  The general Lorentz transformations acting on fermions include two discrete operations: the reversal of time, and parity transformation. Under time reversal we have ${\bf T}\,e_{00} = - e_{00}$ and 
${\bf T}\,{\rm det}(e)= -{\rm det}(e)$, and under parity transformation --  ${\bf P}\,e_{00} = e_{00}$ and 
${\bf P}\,{\rm det}(e) =-{\rm det}(e)$. Correspondingly, the fermionic vacuum has the four-fold degeneracy.

In condensed matter the analog of parity transformation takes place in  a topological Lifshitz transition, when the chiral vacuum with Weyl nodes  in the polar distorted superfluid 
$^3$He-A\cite{Askhadullin2014,Dmitriev2015} crosses the vacuum state of the polar phase with a degenerate fermionic tetrad, ${\rm det}(e)=0$.\cite{NissinenVolovik2018} In this transition from "spacetime" to "antispacetime", the chirality of Weyl fermions changes: the left-handed fermions living in the spacetime  transform to the right-handed fermions in "antispacetime".  This transition experiences the nonanalytic behavior of the action at the crossing point. Here we discuss the similar transition from "spacetime to antispacetime" by the time reversal and show that this transition may have analytical properties suggested in Refs. \onlinecite{Diakonov2011,Diakonov2012,Rovelli2012a}.

\section{Negative lapse function and negative temperature}

Let us assume that the lapse function $N({\bf r})$ is the analytical function of the tetrad field. Then  instead of of the conventional Tolman law, $T({\bf r})=T_0/|N({\bf r})|$ in Eq.(\ref{T}),  one would have the modified Tolman law
\begin{equation}
T({\bf r})= \frac{T_0}{N({\bf r})} \,\,, \,\,
N({\bf r})= e_{00}({\bf r})= \frac{1}{ e^0_0({\bf r})} \,.
\label{N}
\end{equation}
For negative  $e_{00}({\bf r})$  (but still positive $g_{00}({\bf r})$), the local temperature $T({\bf r})$ of fermions becomes negative. 

\begin{figure}[t]
\includegraphics[width=\linewidth]{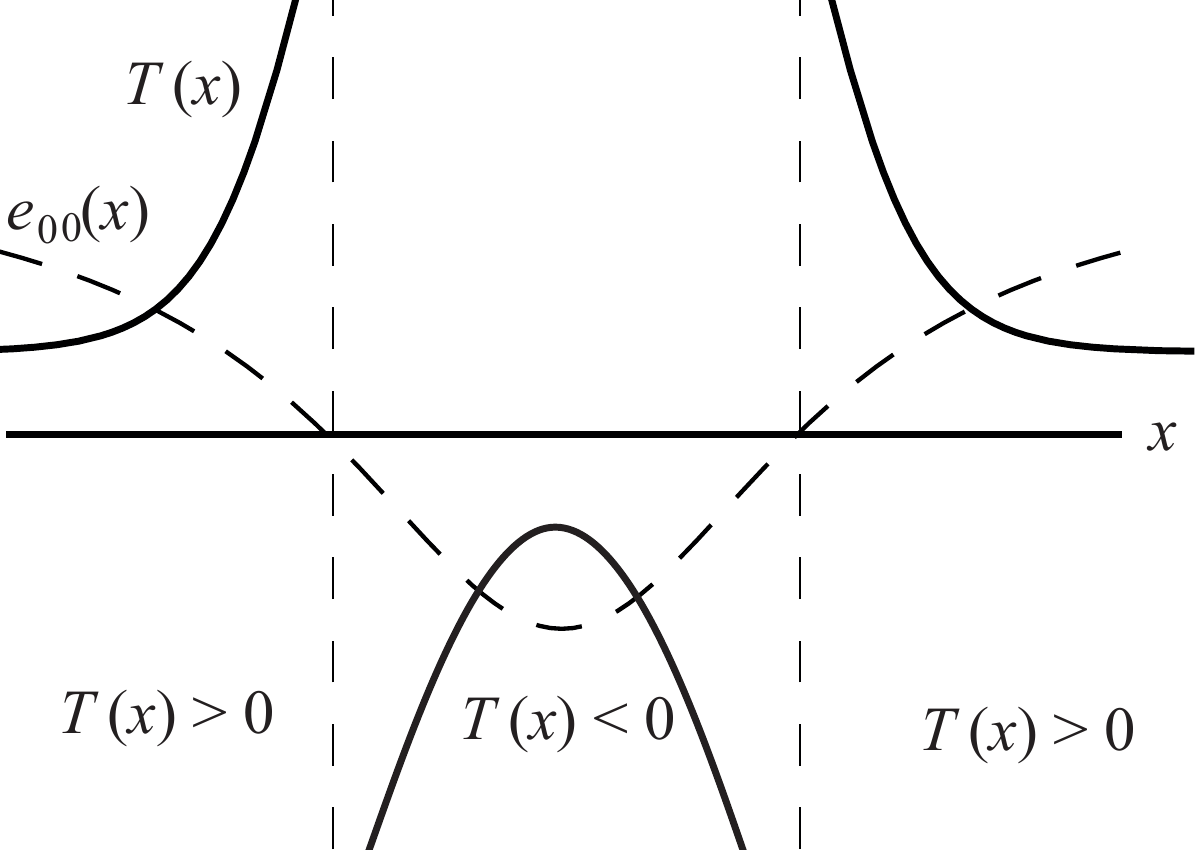}
\caption{ Island of negative lapse function, $N(x)= e_{00}(x)<0$, where the metastable state with negative local temperature is formed according to the modified Tolman law, $T(x)=  T_0/ e_{00}(x) <0$. The global Tolman temperature $T_0$ is constant in space, $T_0= {\rm const}>0$. It is the temperature at infinity, where $e_{00}(\pm \infty)=1$. In this scenario, $e_{00}(x)$ crosses zero, while and temperature $T(x)$ crosses infinity.  The negative temperature state in the island is nonequilibrium, and finally it relaxes to the equilibrium state in Fig. \ref{island2_FigEq} with positive temperature obeying the conventional Tolman law, $T(x)=  T_0/ \vert e_{00}(x)\vert >0$.
}
\label{island2_Fig}
\end{figure}

In Ref. \onlinecite{Rovelli2012b}, $e_{00}(t)$ crosses zero as function of time, and for fermions this corresponds to time reversal operation ${\bf T}$.  
We consider the case when $e_{00}({\bf r})$ crosses zero in space and becomes negative in some island of space, see Fig. \ref{island2_Fig}. Then if Eq.(\ref{T}) is obeyed, the local temperature $T({\bf r})$ in the island is negative. 

\subsection{Crossing infinite temperature}

\begin{figure}[t]
\includegraphics[width=\linewidth]{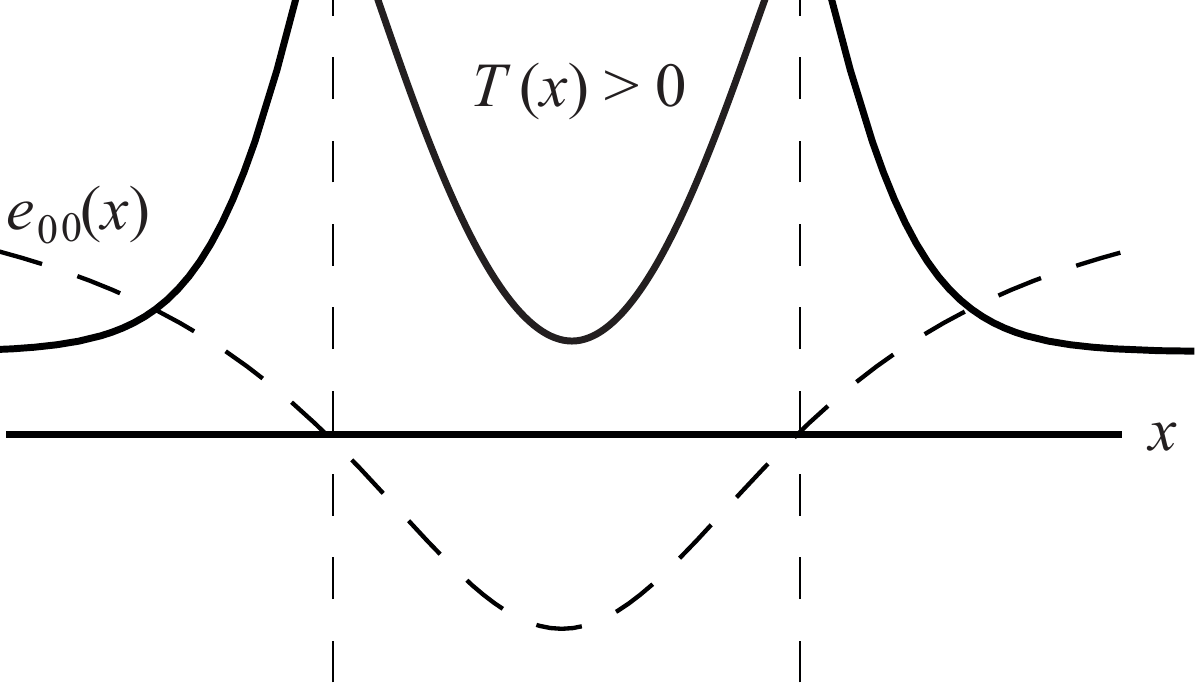}
\caption{ The final stable state of the island of negative lapse function, $N(x)= e_{00}(x)<0$, in Fig. \ref{island2_Fig}. The local temperature obeys the conventional Tolman law, $T(x)=  T_0/ \vert e_{00}(x)\vert >0$, and  is positive everywhere. 
}
\label{island2_FigEq}
\end{figure}

\begin{figure}[t]
\includegraphics[width=\linewidth]{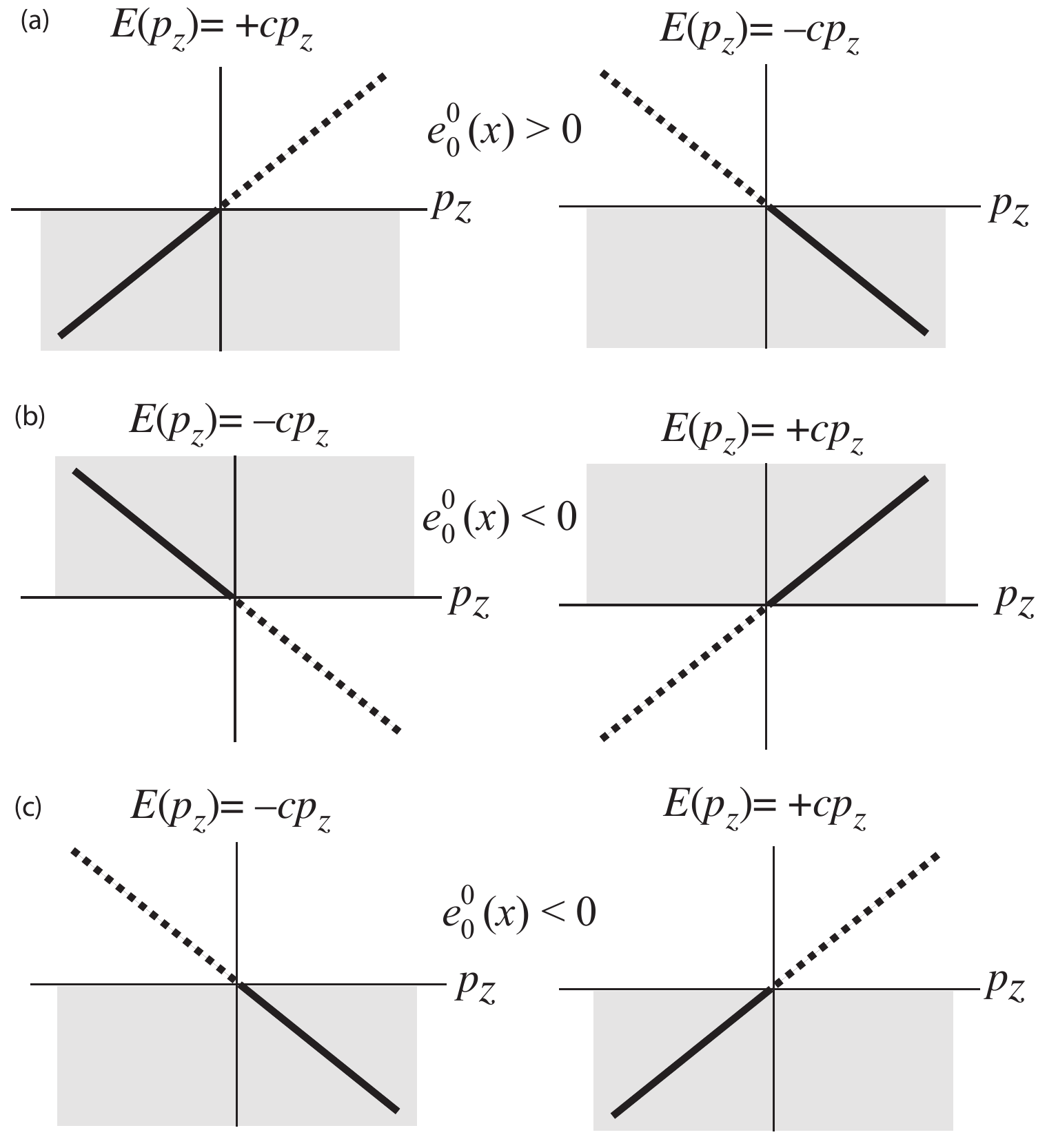}
\caption{ 1D chiral fermions with spectrum $E^2=c^2 p_z^2$. (a) Conventional equilibrium state  at $T({\bf r})=+0$: the negative states are occupied (solid lines), and positive energy states are empty (dashed lines). The entropy in this ground state is $S=0$. (b): Quasiequilibrium state with $S=0$ in the island with opposite $e^0_0<0$. The positive energy states are occupied  (solid lines), and negative energy states are empty (dashed lines). The inverse population of levels corresponds to $T({\bf r})=-0$. (c) The final equilibrium state in the island with opposite $e^0_0<0$ corresponds to $T({\bf r})=+0$.
}
\label{1Dspectrum}
\end{figure}

Formation of negative temperature in the island with negative $e_{00}({\bf r})$ can be explained in the following way. For the fermions, the crossing $e_{00}=0$ corresponds to the change of the Hamiltonian ${\cal H} \rightarrow -{\cal H}$.
 When the island is formed, then immediately after formation one has the state with inverse filling of the particle energy levels, see Fig. \ref{1Dspectrum} for the case of 1+1 relativisitic spectrum. This corresponds to the negative local temperature, $T({\bf r})<0$. 

Though in general the negative temperature state with inverse population is not in full equilibrium,  in principle,  it can be made locally stable,
 see e.g. Ref.\onlinecite{Broun2013}. Anyway, finally the state in the island relaxes to the fully equilibrium state in Fig. \ref{island2_FigEq} with positive temperature, $T(x)=  T_0/ \vert e_{00}(x)\vert 
=  T_0/ \vert N(x)\vert >0$. 

\begin{figure}[t]
\includegraphics[width=\linewidth]{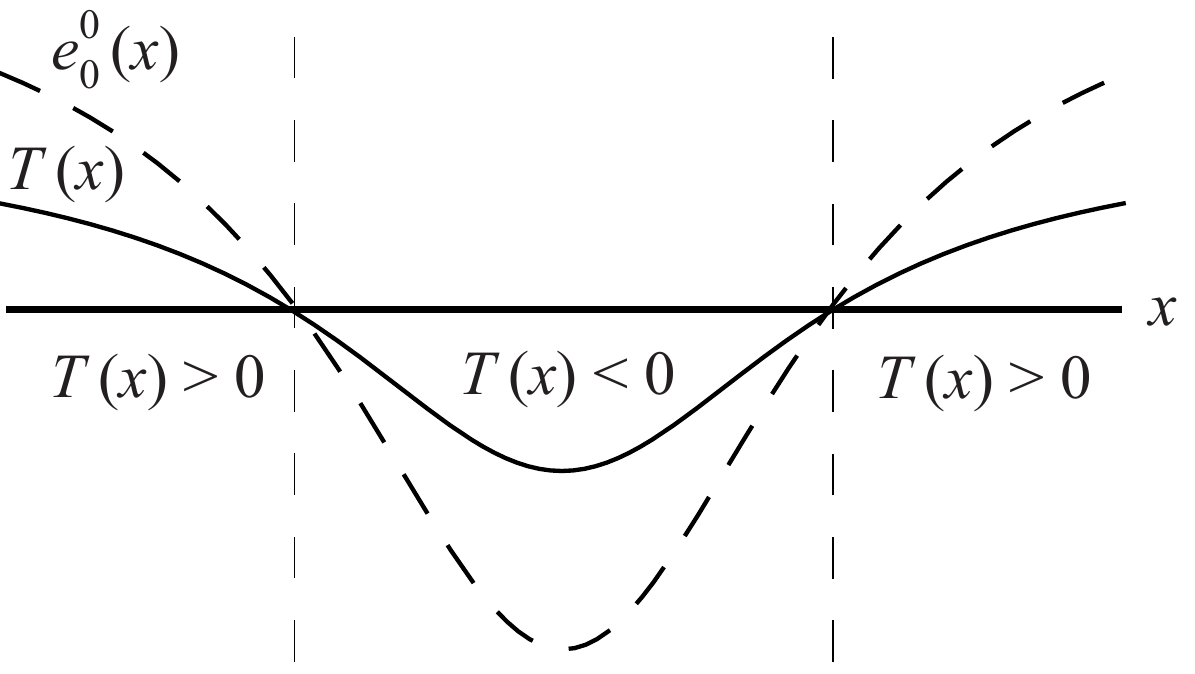}
\caption{ Island of negative lapse function, $N(x)= 1/e^0_0(x)<0$, where the metastable state with negative local temperature is formed, $T(x)= e^0_0(x) T_0 <0$.  In this scenario, both $e^0_0(x)$ and the temperature $T(x)= e^0_0(x)T_0$ cross zero. As before, this state finally relaxes to equilibrium state, where the temperature is everywhere positive.
}
\label{island_Fig}
\end{figure}

In  Fig. \ref{island2_Fig} the  change of sign of $N$ occurs by crossing the boundary with infinite temperature,  i.e. $\beta({\bf r})=1/T({\bf r})$ crosses zero. 
We assume the band structure of the fermionic vacuum, i.e. a finite energy cut-off to avoid divergencies. We also assume that at spatial infinity one has the equilibrium vacuum state with $e_{00}(\infty)=1$. Of course, one can use at infinity another equilibrium degenerate state of the vacuum with $e_{00}(\infty)=-1$ and correspondingly $e_{00}({\bf r})>0$ in the island. In this case,  instead of Eq.(\ref{N}) one should use the equations $N({\bf r})=- e_{00}({\bf r})$ and $T({\bf r})= -e^0_0({\bf r}) T_0$.

\subsection{Crossing zero temperature}

Fig. \ref{island_Fig} demonstrates the case, when $e^0_0({\bf r})$ crosses zero instead of $e_{00}({\bf r})$. In this case  $T({\bf r})$ changes sign by crossing zero temperature.
Such situation may take place in  Weyl semimetals, where the inverse Green's function in the vicinity of the Weyl point  is:\cite{Volovik2003}
\begin{equation}
 G^{-1}=e^\mu_a \sigma^a\left(k_\mu-qA_\mu \right)\,.
\label{G}
\end{equation}
Here $\sigma^a$ are Pauli matrices with $a=0,1,2,3$; $q A_\mu$ marks the positions of two Weyl points in $k$-space; and $q=\pm 1$ is the effective charge in synthetic electormagnetic field $A_\mu$. The tetrad  element $e^0_0$ may change sign due to interaction between fermions.\cite{NissinenVolovik2017}

\subsection{Modified Tolman law for chemical potential}

\begin{figure}[t]
\includegraphics[width=\linewidth]{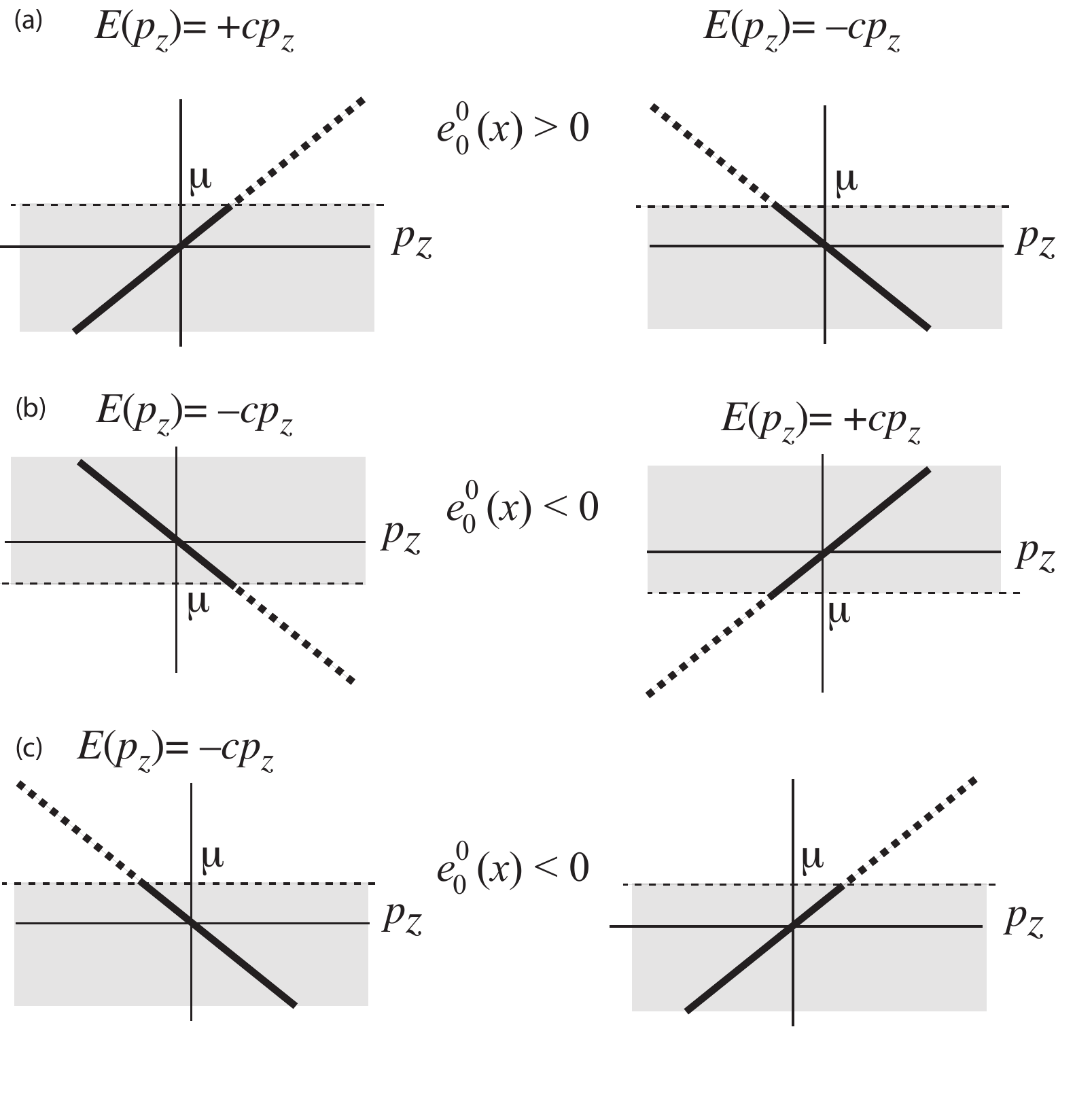}
\caption{(a) The state with positive chemical potential at $x=\pm \infty$. (b) The chemical potential changes sign in the island of negative lapse in the metastable state immediately after formation of the island according to the modified Tolman law $\mu({\bf r})= e^0_0({\bf r})\mu_0$.  (c) The final equilibrium state in the island of negative lapse where $\mu({\bf r})= \vert e^0_0({\bf r})\vert \mu_0$.
}
\label{1Dmu}
\end{figure}

In the discussed approach of Eq.(\ref{N}), the Tolman law  for the chemical potential of relativistic fermions,  $\mu({\bf r})= \mu_0/\sqrt{g_{00}({\bf r})}$,\cite{Balazs1958} is also modified for the metastable state in the island of negative lapse function.  The local chemical potential $\mu({\bf r})= e^0_0({\bf r})\mu_0$ changes sign in the island in the metastable state according to the modified Tolman law, but finally relaxes to the value 
$\mu({\bf r})= \vert e^0_0({\bf r})\vert \mu_0$ obeying the conventional Tolman law,
see Fig.\ref{1Dmu},

\section{Discussion}

So, while the dynamics in the negative lapse region  may correspond to the inverse arrow of time for fermions,\cite{Rovelli2012b}  the thermodynamics in this region corresponds to the negative temperature.
For 3+1d massless fermions, the thermodynamic energy density and the entropy density, assuming that far from the island there is Minkowski spacetime with $e^0_0(\infty)=1$, are:
\begin{equation}
\epsilon({\bf r}) = {\rm sign}(e^0_0({\bf r})) \frac{7\pi^2}{120} T^4({\bf r})\,,
\label{E}
\end{equation}
\begin{equation}
s({\bf r}) = {\rm sign}(e^0_0({\bf r})) \frac{7\pi^2}{90} T^3({\bf r}) =\frac{7\pi^2}{90} T_0^3|e^0_0({\bf r)}|^3>0\,.
\label{s}
\end{equation}

For bosons, the positive and negative lapse functions are intistinguishable. However, if bosons are composite, i.e. made of fermions, they may inherit  from the fermions the negative $T({\bf r})$ in the island, and thus they may distinguish the antispacetimes in the island via thermodynamics. The state with negative temperature in the island is unstable, and finally relaxes to the equilibrium state. That is why the metastability will be noticed both by fermions and bosons.  In the final equilibrium state, the spacetime and antispacetime can be also resolved, since in Standard Model the left-handed and right-handed fermions belong to different representations of the $SU(2)$ group. 

As is demonstrated in Ref. \onlinecite{NissinenVolovik2018} on example of the Weyl superfluid, the action in terms of tetrads is non-analytic. For example, the action for the effective gauge field is shown to be proportional to  $\sqrt{-g}=|{\rm det}(e)|$. This is contrary to the action proportional to ${\rm det}(e)$, which has been suggested in Refs. \onlinecite{Diakonov2011,Diakonov2012,Rovelli2012a}. The nonanalytic behavior of the action takes place when the boundary is crossed between two equilibrium degenerate states with different signs of ${\rm det}(e)$ or $N({\bf r})$. In both equilibrium states the temperature is positive, and the Tolman law is given by the nonanalytic equation (\ref{T}). The analytic action suggested in Refs. \onlinecite{Diakonov2011,Diakonov2012,Rovelli2012a} can be restored in the case considered here, when the boundary is crossed between the  equilibrium state with positive lapse  $N({\bf r})>0$ and  the   nonequilibrium state in the island  with negative lapse  $N({\bf r})<0$ and negative $T({\bf r})<0$. In this case the Tolman law is the analytic function of $N({\bf r})$ in Eq.(\ref{N}), as well as the action.

In Ref. \onlinecite{Rovelli2012b}  three alternatives to the problem of antispacetime were suggested: 
(i) Antispacetime does not
exist, and ${\rm det} \,(e)>0$ should constrain the gravity path integral;
(ii) Antispacetime exists, but the action depends on $|{\rm det} \,(e)|$, rather than on  ${\rm det} \,(e)$. (iii) Antispacetime exists and contributes nontrivially to quantum gravity. 

Our consideration suggests that the antispacetime may exist with two possible realizations. The option (ii) takes place in case of full equilibrium both in spacetime and in antispacetime, and in this case the action is non-analytic \cite{NissinenVolovik2018} in the same way as the conventional Tolman law. The option (iii) with the analytic behavior of the action takes place in the quasiequilibrium state with the negative temperature in the island, which is formed immediately after formation of the island. However, after relaxation to the full equilibrium the nonanalytic behavior of the action and of the Tolman law is restored.

Two remarks are in order.
First, let us note that the realization of the hypersurfaces, at which ${\rm det} \,(e)$ crosses zero or infinity, may require consideration beyond the Einstein general relativity. The similar problem arises for the 
 hypersurfaces, at which the Newton constant changes sign.\cite{Starobinsky1981}

Second, let us mention that Eq.(\ref{G}) for the Weyl fermions can be written in a form, which does not contain 
the Planck constant $\hbar$:
\begin{equation}
 e^\mu_a \sigma^a\left(\partial_\mu-qA_\mu \right)\Psi=0\,.
\label{G2}
\end{equation}
 Of course, if $\hbar$ is a fundamental constant, one can choose units in which $\hbar=1$. But in a similar manner as Weyl fermions, gauge fields and tetrad gravity emerge in 
the vicinity of the Weyl point,\cite{FrogNielBook,Volovik2003,Horava2005,VolovikZubkov2014}  the $\hbar$ can also be the emergent quantity rather than the fundamental constant. In particular, it can be expressed in terms of tetrads. In the  Schr\"odinger equation for massive particles, which is obtained as the nonrelativistic limit of Dirac equation with tetrad fields, the Planck constant  emerges as the value of $e^0_0$ in the Minkowski vacuum state (see Eq.(45) in Ref. \onlinecite{Volovik2009}  or Eq.(5.3) in the 
preprint version of  Ref. \onlinecite{Volovik2009}):
\begin{equation}
 \hbar\equiv \sqrt{-g^{00}(\infty)}=e^0_0 (\infty)\,.
\label{hbar}
\end{equation}
In this case the transition to antispacetime is accompanied by a change in the sign of the Planck constant.

{\bf Acknowledgements}. 
I thank Jaakko Nissinen for discussions. This work has been supported by the European Research Council (ERC) under the European Union's Horizon 2020 research and innovation programme (Grant Agreement No. 694248).


\begin{thebibliography}{99}

\bibitem{Rovelli2012b}
M. Christodoulou, A. Riello, C. Rovelli,
How to detect an anti-spacetime,
Int. J. Mod. Phys. D {\bf 21}, 1242014 (2012),
arXiv:1206.3903.

\bibitem{Tolman1934}
R. Tolman, 
 {\it Relativity, Thermodynamics, and Cosmology},
Oxford University Press, Oxford (1934).

\bibitem{Visser2018}
J, Santiago and M. Visser,
Gravity's universality: The physics underlying Tolman temperature gradients,
arXiv:1805.05583.


\bibitem{Volovik2003}
G.E. Volovik,
{\it The Universe in a Helium Droplet},
Clarendon Press,  Oxford (2003).

\bibitem{Askhadullin2014}
R.Sh. Askhadullin, V.V. Dmitriev, P.N. Martynov, A.A. Osipov, A. A. Senin, A.N. Yudin,
Anisotropic 2D Larkin-Imry-Ma state in polar distorted ABM phase of $^3$He in ‘nematically ordered’ aerogel,
 JETP Lett. {\bf 100}, 662--668 (2014).

\bibitem{Dmitriev2015}
V. V. Dmitriev, A. A. Senin, A. A. Soldatov, and A. N. Yudin,
Polar phase of superfluid  $^3$He in anisotropic aerogel,
Phys. Rev. Lett. {\bf 115}, 165304 (2015).

\bibitem{NissinenVolovik2018}
J. Nissinen and G.E. Volovik,
Dimensional crossover of effective orbital dynamics in polar distorted  $^3$He-A: Transitions to anti-spacetime,
Phys. Rev. D {\bf 97}, 025018  (2018),

\bibitem{Diakonov2011}
D. Diakonov,
Towards lattice-regularized Quantum Gravity,
  arXiv:1109.0091.

\bibitem{Diakonov2012}
A.A. Vladimirov, D. Diakonov,
Phase transitions in spinor quantum gravity on a lattice,
Phys. Rev. D {\bf 86}, 104019 (2012). 

\bibitem{Rovelli2012a}
C. Rovelli, E. Wilson-Ewing,
 Discrete symmetries in covariant LQG,
Phys. Rev. D {\bf 86}, 064002 (2012),
 arXiv:1205.0733.

\bibitem{Broun2013}
S. Braun, J. P. Ronzheimer, M. Schreiber, S. S. Hodgman, T. Rom, I. Bloch, U. Schneider,
Negative absolute temperature for motional degrees of freedom,
Science {\bf 339},  52--55 (2013).

\bibitem{NissinenVolovik2017}
J. Nissinen and G.E. Volovik,
Type-III and IV interacting Weyl points,
Pisma ZhETF {\bf 105}, 442--443 (2017);
JETP Lett.  {\bf 105},  447--452 (2017),
arXiv:1702.04624.

\bibitem{Balazs1958}
N.L. Balazs,  
On relativistic thermodynamics,
Astrophysical Journal {\bf 128}, 398--405 (1958).

\bibitem{Starobinsky1981}
A. A. Starobinsky, 
Can the effective gravitational constant become negative?
Sov. Astron. Lett.  {\bf 7}, 36 (1981); Pisma v Astron. Zh. {\bf 7}, 67--72, 1981).

\bibitem{FrogNielBook}
C.D. Froggatt  and  H.B. Nielsen,
 {\it Origin of Symmetry}, World Scientific, Singapore (1991).

\bibitem{Horava2005}  
P. Ho\v{r}ava,
Stability of Fermi surfaces and $K$-theory,
Phys. Rev. Lett. \textbf{95}, 016405 (2005).

\bibitem{VolovikZubkov2014}
G.E. Volovik and M.A. Zubkov,
Emergent Weyl spinors in multi-fermion systems,
Nuclear Physics B {\bf 881}, 514--538  (2014).

\bibitem{Volovik2009}
 G.E. Volovik,  
$\hbar$ as parameter of Minkowski metric in effective theory,
JETP Lett. {\bf 90}, 697--704 (2009);
arXiv:0904.1965.

\end{thebibliography}
\end{document}